\documentclass[conference]{IEEEtran}
\IEEEoverridecommandlockouts
\usepackage{cite}
\usepackage{amsmath,amssymb,amsfonts}
\usepackage{algorithmic}
\usepackage[linesnumbered]{algorithm2e}
\usepackage{graphicx}
\usepackage{svg}       
\usepackage{textcomp}
\usepackage{xcolor}
\usepackage{comment}
\usepackage{booktabs}
\usepackage{todonotes}
\usepackage{placeins}
\usepackage{tablefootnote}
\RestyleAlgo{ruled}
\SetKwComment{Comment}{/* }{ */}

\def\BibTeX{{\rm B\kern-.05em{\sc i\kern-.025em b}\kern-.08em
    T\kern-.1667em\lower.7ex\hbox{E}\kern-.125emX}}
\begin{document}

\title{Efficient Hardware Implementation of Constant Time Sampling for HQC
\\
}

\author{\IEEEauthorblockN{Maximilian Schöffel}
\IEEEauthorblockA{\textit{Microelectronic Design Research Group} \\
\textit{University of Kaiserslautern-Landau}\\
 Kaiserslautern, Germany \\
 m.schoeffel@rptu.de}
 \and
\IEEEauthorblockN{Johannes Feldmann}
\IEEEauthorblockA{\textit{Microelectronic Design Research Group} \\
\textit{University of Kaiserslautern-Landau}\\
Kaiserslautern, Germany \\
j.feldmann@rptu.de}
\and
\IEEEauthorblockN{Norbert Wehn}
\IEEEauthorblockA{\textit{Microelectronic Design Research Group} \\
\textit{University of Kaiserslautern-Landau}\\
 Kaiserslautern, Germany \\
 norbert.wehn@rptu.de}
}

\maketitle

\newcommand\copyrighttext{%
  \footnotesize We acknowledge that the key idea of this paper (performing the uniqueness check of the polynomials in explicit rather than support representation) was published in \textit{Fast and Efficient Hardware Implementation of HQC}, Appendix 1.B,  by Deshpande et al. during the review process of this paper.
  Therefore, we do not further pursue a conference publication.
  }
  
\newcommand\copyrightnotice{%
\begin{tikzpicture}[remember picture,overlay]
\node[anchor=south,yshift=10pt] at (current page.south) {\fbox{\parbox{\dimexpr\textwidth-\fboxsep-\fboxrule\relax}{\copyrighttext}}};
\end{tikzpicture}%
}
    
\begin{abstract}
HQC is one of the code-based finalists in the last round of the NIST post quantum cryptography standardization process.
In this process, security and implementation efficiency are key metrics for the selection of the candidates.
A critical compute kernel with respect to efficient hardware implementations and security in HQC is the sampling method used to derive random numbers.
Due to its security criticality, recently an updated sampling algorithm was presented to increase its robustness against side-channel attacks.

In this paper, we pursue a cross layer approach to optimize this new sampling algorithm to enable an efficient hardware implementation without comprising the original algorithmic security and side-channel attack robustness.

We compare our cross layer based implementation to a direct hardware implementation of the original algorithm and to optimized implementations of the previous sampler version. 
All implementations are evaluated using the Xilinx Artix 7 FPGA.
Our results show that our approach reduces the latency by a factor of 24 compared to the original algorithm and by a factor of 28 compared to the previously used sampler with significantly less resources.

\end{abstract}


\copyrightnotice

\section{Introduction}
Quantum computers are expected to revolutionize sectors such as medicine, materials science, and artificial intelligence once they reach maturity with adequate computing power.
However, they also pose serious threats on communication security, and it is expected that it is possible to break State of the Art (SoA) public key cryptography by the end of this decade~\cite{mosca2018}.
Therefore, the United States National Institute for Standards and Technology (US NIST) is currently conducting a process to find new, quantum computer resistant cryptographic algorithms (Post-Quantum Cryptography or PQC)~\cite{nist2023}.

Hamming Quasi Cyclic (HQC)~\cite{melchor2021hamming} is one of the code-based candidates for standardization that has advanced to the final round of the NIST PQC process.
Compared to the already standardized lattice-based algorithm KYBER, implementations of code-based algorithms have both a larger computational complexity and larger memory footprint.
However, viable alternatives to lattice-based algorithms are already required today in case that the relatively new lattice-based algorithms turn out to be insecure in future (Crypto Agility).
As a result, the application requirements in environments such as the Industrial Internet of Things (IIoT), which are limited by computing power and available energy but have strict timing constraints, can often only be met by using dedicated hardware accelerators.

For such applications, several papers on hardware implementations have already been published.
The authors of HQC proposed a High-Level Synthesis (HLS)-based design that outperforms the remaining code-based candidates in the final round of the NIST process~\cite{aguilar2022towards}.
A HW/SW co-design of HQC that targets IoT applications was proposed in~\cite{schoffel2022code}, and the authors found that the memory and the sampling unit are the main contributors to the area requirement.
Furthermore, they showed that sampling and polynomial multiplication are the main drivers of the HQC latency.
One of these bottlenecks, the polynomial multiplication, was addressed by the LEAP multiplier in~\cite{tu2023leap}.

Secure and efficient sampling algorithms to derive random numbers are still a subject of research, and a detailed study of different algorithms and protection measures with respect to power side channels was conducted in~\cite{krausz2023holistic}.
While their focus was on algorithms rather than hardware implementations which ensure robustness against power SCAs, the authors also provided a hardware implementation, which, however, requires significant computation time and was designed primarily for BIKE, the other code-based candidate in the NIST process.

Recently, the authors of HQC introduced an updated, more secure sampling procedure based on Sendrier's modified Fisher-Yates approach~\cite{sendrier2021secure} as a response to successful timing SCAs on the HQC sampling procedure~\cite{guo2022don}.

In conclusion, the focus of research so far has been mainly on polynomial ring multiplication.
But for the other critical factor -- the new sampling algorithm -- an efficient solution has yet to be explored. 
Existing implementations are either prone to timing SCAs or have high computation time.

In this paper, we present to the best of our knowledge the first cross-layer approach for this new sampler.
We investigate how the interrelationships between the algorithmic layer, which provides the base for a timing SCA-resistant implementation, and the hardware implementation layer can be exploited to maximize implementation efficiency without comprising security.
Our new contributions are:
\begin{enumerate}
    \item A reduction of the computational complexity from $\mathcal{O}(n^2)$ to $\mathcal{O}(n)$ through a new approach to represent the polynomials during the sampling procedure.
    \item A hardware implementation with a pipelining scheme that is robust against timing side channel attacks. The resource efficiency (latency / number of required FPGA resources) is increased by jointly considering the implementation of the sampling procedure and polynomial ring arithmetic.
    \item A detailed comparison of implementation results of a standard implementation, our cross layer approach and optimized implementations of the previous sampler on a Xilinx Artix 7 FPGA.
\end{enumerate}

Our new algorithm requires an update of the original HQC specification, as the random numbers derived from the same seed are not equal to the original algorithm.
However, since standardization is still in progress, we encourage to adopt these changes as they do not add complexity to software implementations nor reduce security.

This paper is structured as follows.
In Section II, a background of HQC and the sampling algorithm will be provided.
In Section III, the algorithmic improvements will be explained.
In Section IV, the hardware design will be presented.
In Section V, the results of the hardware implementation will be provided, evaluated and compared with the SoA.
In Section VI, we draw a conclusion.

\section{Background}
HQC~\cite{melchor2021hamming} is a Key-Encapsulation Mechanism (KEM) and its security is based on the hardness of the syndrome decoding problem.
Its design rational bases on concatenated Reed-Muller/Reed-Solomon code $\mathcal{C}$ and erroneous codewords are generated by adding and multiplying random values on the secret message to hide the same.
These arithmetic operations are performed on sparse (low hamming weight) and dense (high hamming weight) polynomials $v$ in the ring $\mathcal{R}=\mathbb{F}_2[X]/(X^n-1)$, and $v$ can be represented in two different ways:

\begin{enumerate}
    \item \textbf{Explicit representation:} A bit array $v$ of length $n=17669$ where each bit entry $v_i$ represents the coefficient $v_i$ in $v=\sum_{i=0}^{n-1} v_i \cdot x^i$.
    The explicit representation is used for dense polynomials.
    \item \textbf{Support representation:} An array $c$ of length $\omega \leq 75$, where each integer entry $c_i$ represents the coordinate of a non-zero coefficient in the corresponding $v$: $v=\sum_{i=0}^{\omega-1} x^{c_i}$. 
    The support representation is used for sparse polynomials.
\end{enumerate}

HQC is available in three different parameter sets, in this work we refer to HQC-128.
Besides the arithmetic in $\mathcal{R}$, the major contributor to the computational complexity of HQC is the sampling procedure~\cite{schoffel2022code}, where random error polynomials are generated through the expansion of a secretly, truly random generated seed using the Extendable Output Function (XOF) SHAKE~\cite{nist2015}.
For HQC to operate securely and correctly, the sampled error polynomials must fulfill two criteria.
First, as the sampling procedure directly involves the computation of security critical values, a constant execution time (independent from the sampled values) is a crucial countermeasure against SCAs.
In addition, the sampled errors must meet strict requirements regarding their Hamming weight $\omega$, since a too low Hamming weight would allow attacks on the KEM, and a too large Hamming weight would mean that the original message could not be decrypted by the communication partners as it would exceed the error correction capability of $\mathcal{C}$.
Therefore, Algorithm~\ref{orig_algo} has been proposed by the HQC authors based on the previous works of Sendrier~\cite{sendrier2021secure} to meet both requirements.

The algorithm operates with the error polynomials in their support form and, if required, converts them to explicit representation once the sampling is completed (Line 12).
An array with $\omega$ random words is generated through the expansion of $seed$ with SHAKE in Line 1.
This $randomwords$ array is further processed in the $mod\_loop$, which implements the "sampling with bias" as introduced by Sendrier~\cite{sendrier2021secure}.
In $unique\_check\_loop$, each of the sampled values in $support$ are compared with all previously sampled values.
If there is a duplicated value, the $found$ flag is set and $support[i]$ receives the loop iterator $i$ as a value instead.
Because of the biased sample in Line 3 and the fact that $unique\_check\_loop$ iterates backwards, the condition $support[i] \geq i$ is always satisfied.
These measures guarantee that the hamming weight of the resulting polynomial is exactly $\omega$.

\begin{algorithm}
\label{orig_algo}
\caption{Constant weight sampling algorithm based on~\cite{sendrier2021secure} with modifications by the authors of HQC. Note that $word$ refers to 32-bit values. The "$\xleftarrow[]{\text{\$}}$" operator refers to sampling bytes from a random distribution.}
\KwData{$n=17669, seed, \omega$}
\KwResult{$v \in \mathcal{R}=\mathbb{F}_2[X]/(X^n-1)$ with hamming weight $\omega$}
$randomwords \xleftarrow[]{\text{\$}} prng(seed,\omega)$\;
mod\_loop: \For{$0 \leq i \leq$ $\omega - 1$}{
    $support[i] \gets i + (randomwords[i]\mod(n-i))$\;
}
unique\_check\_loop: \For{ $(\omega-1)\geq i\geq0$}{
    $found \gets 0$\;
    \For{$(i+1) \leq j \leq$ $\omega-1$}{
        $found \gets compare(support[i],support[j])$\;
    }
    $support[i] \gets found $ $ ?$ $ i : support[i]$\;
}
$v \gets transform(support)$\;
\Return v

\end{algorithm}

In general, sampling and ring arithmetic in HQC are used in the following sequences:

\begin{equation}
\label{eq:sample_h}
h \xleftarrow[]{\text{\$}} \mathcal{R}, \\
\end{equation}
\begin{equation}
\label{eq:sample_xy}
    (x,y)\xleftarrow[]{\text{\$}} \mathcal{R}^2, \\
\end{equation}
\begin{equation}
\label{eq:r_arithmetic}
    z \gets h \cdot y+ x,
\end{equation}

where $h$ is a dense polynomial which is directly sampled through seed expansion by using SHAKE, and $x$ and $y$ are sparse (error) polynomials with hamming weight $\omega$ sampled with Algorithm~\ref{orig_algo}.

\section{Algorithmic Optimizations}
\label{ch:cross}
In the following, we explain the algorithmic optimizations in our cross layer approach and evaluate each one in terms of its security impact, compatibility with the original algorithm, effect on hardware implementations, and drawbacks.

\subsection{Reducing computational complexity: Uniqueness Check}
\label{sec:unique}
\textbf{Description:}
The uniqueness check during the sampling procedure ensures that each of the sampled values occurs only once, thus guaranteeing the Hamming weight of the result.
In the support representation in Algorithm~\ref{orig_algo}, each element of the array holds the coordinate of a "1" coefficient in the polynomial.
To ensure that the same coordinate only occurs once in the array, a comparison with all previously sampled coordinates in the array is required, as all of them could hold the same coordinate.
This causes the computational complexity of the algorithm to be $\mathcal{O}(\omega^2)$.

\textbf{Improvement:}
We propose to store the sampled values in explicit representation instead.
There, a single comparison with the currently stored coefficient bit $v_i$ at the sampled coordinate $c_i$ is necessary to determine the uniqueness of this value, thus significantly reducing the complexity from $\mathcal{O}(\omega^2)$ to $\mathcal{O}(\omega)$.

\textbf{Security:}
On our target platform (FPGA), the polynomials are stored in Block Random Access Memory (BRAM) or Look Up Tables (LUTs) without any intermediate cache, thus guaranteeing a memory address independent access latency.
On other platforms, this measure can induce side channel attack possibilities depending on the hardware architecture.
For architectures with address dependent memory access latency secret information (in this case the location of the word inside the corresponding array) can be leaked.
This is for example the case for software implementations on cache-based computer architectures.

\textbf{Hardware:}
The major drawback of this method is that it increases the memory requirement for the polynomial by a factor of 15.7 (17669 bits instead of 1125 bits), thus memory has to be traded-off against runtime.
Furthermore, for the subsequent polynomial multiplication in $\mathcal{R}$ to be efficient, one of the polynomials ($y$, see Equation~\ref{eq:r_arithmetic}) needs to be available in support representation.

\textbf{Compatibility:}
This optimization is bit-true, thus it maintains seed compatibility.

\subsection{Increasing Throughput: Enabling Efficient Pipelining}
\label{sec:pipeline}
\textbf{Description:}
The algorithm can be further improved to enable an efficient pipelining scheme that increases the hardware throughput, i.e. number of coordinates sampled per time.
In Algorithm~\ref{orig_algo}, the \texttt{mod\_loop} requires the first element, $randomwords[0]$, to calculate $support[0]$, whereas the \texttt{unique\_check\_loop} requires $support[\omega-1]$ in its first iteration, thus introducing a data dependency between both loops which prohibits efficient pipelining.

\textbf{Improvement:}
 We suggest an implicit inversion of the $randomwords$ array, where the first word returned by $prng$ is used as $randomwords[\omega-1]$ instead of $randomwords[0]$.
Using this approach, the \texttt{mod\_loop} and \texttt{unique\_check\_loop} loops can be combined into a single loop, allowing efficient pipelining.
Since SHAKE (the prng algorithm) returns random values sequentially, it is possible to operate directly on the returned word in the pipelined loop instead of waiting for the random array to be returned as a whole, as described in our improved Algorithm~\ref{modified_algo}.

\textbf{Security:}
Given that SHAKE is a cryptographically secure XOF~\cite{nist2015}, it holds that $prng$ returns uniformly randomly distributed words.
Hence, from a security perspective, the quality of the content in $randomwords[i]$ and $randomwords[i+1]$ is equivalent $\forall i$ in the given context. \

\textbf{Hardware:}
Pipelining and concurrent execution can significantly reduce the time it takes to sample a polynomial.
In addition, including random words and support computation in the same pipeline loop eliminates the BRAM required to store these arrays as a whole.

\textbf{Compatibility:}
Seed compatibility is not given since the same seed yield to different sampled polynomials compared to the reference implementation supplied by the authors of HQC.
Therefore, it requires an update of the original algorithmic specification, e.g. by replacing $randomwords[i]$ in Line 3 in Algorithm~\ref{orig_algo} with $randomwords[\omega-1-i]$.

\begin{algorithm}
\label{modified_algo}
\caption{Our proposal for the sampling algorithm}
\KwData{$n=17669, seed, \omega$}
\KwResult{$v \in \mathcal{R}=\mathbb{F}_2[X]/(X^n-1)$ with hamming weight $\omega$}
\For{ $(\omega-1)\geq i\geq0$}{
    $randomwords[i] \xleftarrow[]{\text{\$}} prng(seed,1)$\;
    $support[i] \gets i + (randomwords[i]\mod(n-i))$\;
    \If{$v[support[i]] = 1$} {
        $v[i] = 1$\;
    }
    \Else
    {
         $v[support[i]] = 1$\;
    }
}
\Return v

\end{algorithm}

\subsection{Decreasing memory footprint: Joint consideration of the sampling and arithmetic in $\mathcal{R}=\mathbb{F}_2[X]/(X^n-1)$}
\label{sec:joint}
\textbf{Description:}
In SoA implementations, sampling and the arithmetic in $\mathcal{R}$ are treated as separated functions without considering synergies between both to increase their efficiency (number of hardware resources and latency).
Memories are used in the following way for these two functions (e.g.~\cite{aguilar2022towards}~\cite{schoffel2022code}):

\begin{enumerate}
    \item $h$ is sampled and stored in the Random Access Memory (RAM) ($RAM0$) in explicit representation.
    \item $x$ is sampled and stored in $RAM1$ in explicit representation.
    \item $y$ is sampled and stored in $RAM2$ in support representation.
    \item $z'$, the product of h and y, is computed and stored in $RAM3$. 
    The product of $h$ and $y$ requires twice the storage of the explicit representation.
    \item $z$ is calculated by the modular reduction of $z'$ and adding $x$.
\end{enumerate}

Thus, four RAMs are required to execute this procedure.
Notably, memory has been shown to be the major contributor to the area requirement of the HQC implementation~\cite{schoffel2022code}. \

\textbf{Improvement:} We propose the following procedure instead:

\begin{enumerate}
    \item $x$ is sampled and stored in $RAM0$ in explicit representation.
    \item $y$ is sampled and stored in $RAM1$ in explicit representation during the uniqueness check, and concurrently stored in $RAM2$ in support representation. The support representation of $y$ is later required for the dense-sparse polynomial multiplication.
    \item $h$ is sampled and stored in $RAM1$, thus overwriting the no longer required explicit representation of $y$.
    \item $z$ is computed and stored in $RAM0$. Each intermediate result of $z$ is reduced modulo $X^n-1$, thereby only requiring storage for single explicit representation.
\end{enumerate} \

\textbf{Hardware:}
This procedure has four advantages:
\begin{enumerate}
    \item It eliminates the need for an additional RAM when sampling the sparse polynomial $y$ in explicit representation (see Section ~\ref{sec:unique}), since $RAM1$ can be reused for $h$ after sampling $y$.
    \item $x$ can directly be stored inside the result RAM for $z$, thus removing its requirement for an additional RAM.
    \item The required memory size for $z$ is halved.
    \item The schoolbook approach in~\cite{aguilar2022towards} and~\cite{schoffel2022code} multiplies $h \cdot y$ wordwise, i.e. each memory word in $h$ is multiplied by each coordinate in $y$ based on a windowing method.
    The intermediate results are stored in $RAM0$ and repeatedly added to the subsequent intermediate results (which corresponds to an XOR operation since the elements of the polynomial are in $\mathbb{F}_2$).
    This procedure can be used to implicitly add x during the multiplication steps by preloading $x$ instead of zeros into $RAM0$, thus eliminating the need for an additional computation step.
    This requires to use the schoolbook approach instead of the optimized LEAP multiplier~\cite{tu2023leap}.
\end{enumerate}

\textbf{Security:}
This improvement does not affect the security in the hardware implementation.

\section{Hardware Design}
In our cross-layer approach described in Section~\ref{ch:cross}, we improved the algorithmic layer to enable an efficient hardware implementation with respect to
\begin{enumerate}
    \item \textbf{latency} reduction through the optimizations described in Sections~\ref{sec:unique} and~\ref{sec:pipeline}, and
    \item \textbf{hardware resources} reduction through the optimizations described in Sections~\ref{sec:pipeline} and~\ref{sec:joint}.
\end{enumerate}

In this chapter, we present the challenges and solutions of our hardware design which incorporates the above optimizations.

\subsection{Sampler}
The major challenge of a pipeline schedule for Algorithm~\ref{modified_algo} is to ensure that no secret information gets leaked through the execution time.
Here, the two critical operations are the modulo operation in Line 3 and the uniqueness check procedure and its measures.

For fixed modulus, the Barrett Reduction~\cite{becker2021neon} is often used in hardware implementations to ensure constant time calculation and resource efficiency.
This replaces the division with two multiplications, subtractions and shifts using a pre-computed reduction factor $R$ instead, see Algorithm~\ref{barret_algo}.
However, the modulus of the sampling algorithm is not fixed, but is in the range $[n,n-\omega]$, and therefore requires the storage of $\omega$ precomputed 18-bit values of $R$.
For the given range with $\omega=75$, we observed the following dependency, which drastically reduces the required storage from $18\cdot\omega~bits$ to $\omega+ 18~bits $ at the cost of one addition and subtraction:

\begin{equation}
    R_{i-1} = R_{i} - 14 + LUT[i],
\end{equation}
where $LUT$ is a precomputed one-bit array of length $\omega$ and $R_0 = \frac{2^{32}}{n-(\omega-1)}$.

\begin{figure*}[!ht]
	\centering
		\includegraphics[width=0.9\textwidth]{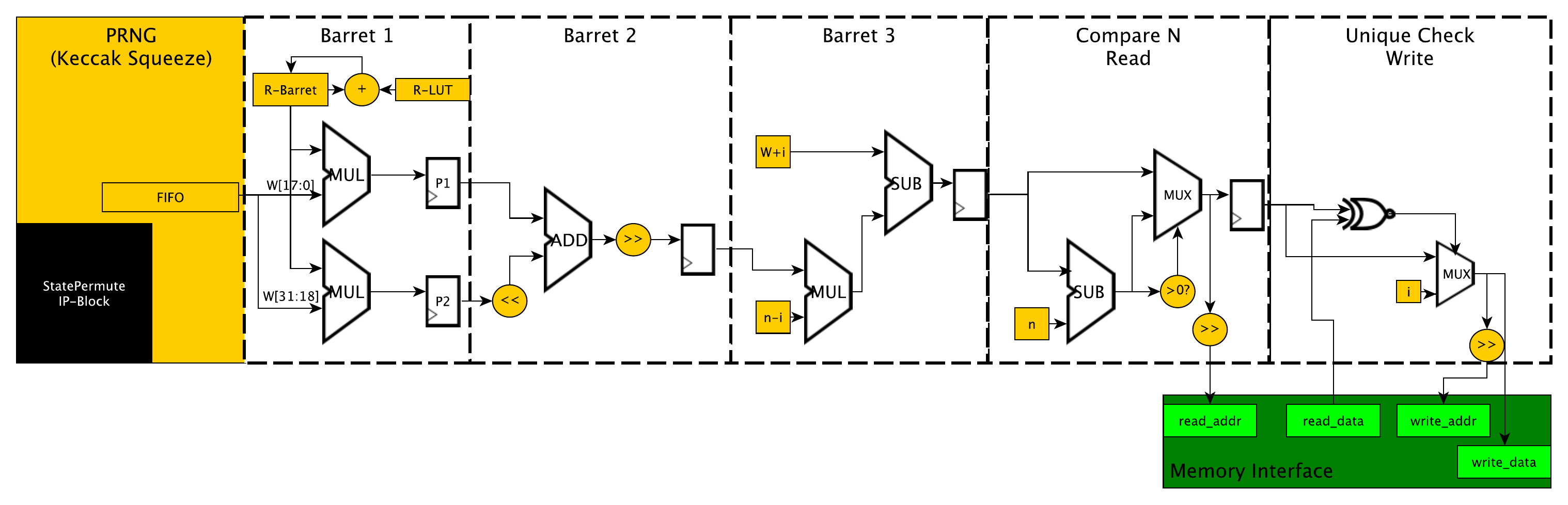}
		\caption{Sketch of the sampling pipeline.}
	\label{fig:pipeline}
\end{figure*}

\begin{algorithm}
\label{barret_algo}
\caption{Barret reduction algorithm used in the hardware implementation based on~\cite{becker2021neon}}
\KwData{$W, M$, $R \gets \lfloor \frac{2^{32}}{M} \rfloor $}
\KwResult{$X= W \mod M$}
$X \gets W - M \cdot ((W \cdot R) >> 32)$\;
\If{$X \geq M$} {
    $X \gets X - M$\;
}
\Return X

\end{algorithm}

For the uniqueness checking procedure, we ensure a constant time implementation as follows.
We assume that the polynomial $v$ is stored in $w$ words of length $m_w$ in a BRAM that allows simultaneous reads and writes, and that one word holds multiple coordinates.
If a new coordinate is to be written, the currently stored data must be read first to prevent overwriting.
In a pipelined architecture, this means that in cycle $i$, reads belonging to the current iteration are performed, and in cycle $i+1$, the corresponding bit in the word is set and written to memory.
This causes two challenges. 
First, a read-after-write hazard could occur if the same address is written and read in the same cycle (i.e., by two successive sampled values), which leads to reading obsolete data in the second access. 
We solve this by detecting and storing the last word written in a register, which is used for the next write if a hazard is detected, instead of the word read from memory.
Second, in most cases, if a non-unique value is found, it requires a write to a different memory word than the one previously read.
However, stalling the pipeline to perform this required read would reveal secret information due to the non-constant execution time.
Therefore, we locally keep record of the memory words belonging to the first $\omega$ coordinates and write them to memory at the end of the sampling procedure.
The actual writing to memory is performed independently of the uniqueness of the respective values, so as not to leak information about the power.

The operation principle of the pipelined sampler is presented in Figure~\ref{fig:pipeline}.
All stages are designed such that they are able to operate in parallel.
The first stage of the pipeline implements the PRNG word generation.
The words are generated by the SHAKE function, which works by permuting and then squeezing the internal state of the Keccak~\cite{nist2015}.
Our design uses an Intellectual Property (IP) core from the authors of Keccak to perform the state permutation~\cite{bert2022}, the squeezing is executed on a 64-bit boundary.
Three pipeline stages are used to calculate the Barret Reduction, as we use a Karazuba based approach to efficiently map the multiplication to the DSP slices of our FPGA.
The last two stages implement the uniqueness check and memory accesses.

\subsection{Arithmetic in $\mathcal{R}$}
The polynomial multiplication in $\mathcal{R}$ is implemented using a similar approach as in~\cite{aguilar2022towards} and~\cite{schoffel2022code}, with modifications to incorporate our proposed improvements such that addition, multiplication and reduction are executed simultaneously.
To reduce the hardware complexity, we do not read and add the intermediate results in the arithmetic module as in the cited works, but instead, we implemented a modify access that XORs the data currently stored in memory with the incoming data in our memory module.
This approach ensures that our hardware has similar performance (resource utilization, frequency and clock cycles) like the LEAP multiplier~\cite{tu2023leap}.

\section{Results}
An objective comparison for hardware implementations in the US NIST PQC process is difficult because the US NIST only specified the Xilinx Artix 7 product family as target platform.
However, the devices within this family still differ significantly in terms of available resources and speed level, and comparing results from different devices (as done several times in the SoA) is not objective.
Therefore, in our comparison, the same device was used as in the original HQC implementation~\cite{aguilar2022towards}, Xilinx Artix xc7a100tftg256-1 FPGA with the Xilinx Vivado 2020.1 software version.

\subsection{Comparison of sampler implementations}

\begin{table}
    \centering
    \caption{Comparison of latency, frequency, time in microseconds and FPGA resources of different implementations for the sampler function. * notes that the design was implemented on a different FPGA.}
    \label{tab:compare_sampler}
    \begin{tabular}{l|cc|ccc}
            \textbf{Impl.} & \textbf{Latency} & \textbf{Frequency} & \multicolumn{3}{c}{\textbf{FPGA Resources}}\\
        & [Cycles] / [us] & [MHz] & LUT & FF & BRAM \\
        \toprule
        HW/SW~\cite{schoffel2022code}* & 3074 / 30.7 & 100 & 925 & 192 & 0\\
        HLS~\cite{aguilar2022towards} & 3726 / 24.35 & 153 & 1472 & 1021 & 1\\
        \midrule
        Original Alg. & 2602 / 20.82 & 125 & 469 & 265 & 0\\
        \textbf{New Alg.} & \textbf{115 / 0.86} & \textbf{133} & \textbf{453} & \textbf{170} & \textbf{0}\\
        \bottomrule
        \end{tabular}
\end{table}

In Table~\ref{tab:compare_sampler} we compare four different implementations to evaluate the efficiency of optimization~\ref{sec:unique} and~\ref{sec:pipeline}.
For all implementations, we considered the hamming weight $\omega=66$ as required during the key generation procedure of HQC128.
\textit{HW/SW} and \textit{HLS} refer to implementations of the previously used sampling algorithm that is susceptible to timing SCAs as described in~\cite{guo2022don}.
\textit{Original Alg.} and \textit{New Alg.} present the results of our implementations of the original, new sampling procedure and our optimized one.
In all implementations, the hardware of the SHAKE module is not included.
However, \textit{HW/SW} used the same IP block for it as the one employed in our design.

As shown in the table, the implementation of the new algorithm requires less clock cycles than the two preceding versions, while increasing the security via constant time implementation.
The reason for this is that the old algorithm requires more random numbers to be sampled with SHAKE due to rejections, and that it operates on a non-optimal 24-bit rather than 32-bit boundary for its random values, which complicates the access pattern for typical 32-bit or 64-bit memories as used in~\cite{schoffel2022code}.
Furthermore, the results in the table show that more than one order of magnitude ($ 24\times$) of latency decrease is achieved through our optimizations~\ref{sec:pipeline} and~\ref{sec:unique} over \textit{Original Alg.}.

\subsection{Key Generation function}
\begin{table}
\centering
    \caption{Comparison of latency in clock cycles and time in microseconds, frequency and FPGA resources of different implementations for the HQC-128 KeyGen function.}
    \label{tab:compare_keygen}
    \begin{tabular}{l|cc|ccc}
            \textbf{Impl.} & {\textbf{Latency}} & \textbf{Freq.} & \multicolumn{3}{c}{\textbf{Resources}}\\
        & [Cycles] / [us] & [MHz] & LUT & FF & BRAM \\
        \toprule
        HLS~\cite{aguilar2022towards} & 40427 / 269.5 & 150 & 11484 & 8798 & 6\\
        $^\vdash$Poly-Mult. & $^\vdash$21418 / 142.79 & 150 & $^\vdash$6175 & $^\vdash$3701 & 0\\
        HW/SW~\cite{schoffel2022code} & 56000 / 560 & 100 & 8000 & 2400 & 3\\
        \textbf{This work} & \textbf{6492 / 46} & \textbf{141} & \textbf{13834} & \textbf{6963} & \textbf{2.5} \\
         $^\vdash$ $\mathcal{R}$-Module & $^\vdash$ 4885 / 34.65 & 141 & $^\vdash$ 2956 & $^\vdash$ 862 & 0 \\
        \midrule
        KYBER~\cite{xing2021compact}& 3800 / 23.75& 160 & 7412 & 4644 & 3\\
        \bottomrule
        \end{tabular}
\end{table}

A conclusive evaluation of the optimizations presented in this paper requires a view of the entire HQC cryptosystem, not just the sampler.
Therefore, we included the joint design with the $\mathcal{R}$ arithmetic module in the implementation of the HQC key generation function by~\cite{aguilar2022towards}.
For the integration, the sampler module has been further modified to support the SHAKE-based seed expansion to sample $h$.
The results of this integration are illustrated in Table~\ref{tab:compare_keygen}.
As presented, our $\mathcal{R}$ module provides a comparable resource efficiency (latency / number of resources) like the LEAP multiplier~\cite{tu2023leap}.
In addition, optimization~\ref{sec:joint} leads to a notable reduction in the amount of memory required, which has previously been shown to be the largest area contributor in the Application Specific Integrated Circuit (ASIC) implementation of HQC~\cite{schoffel2022code}.
When all optimizations are combined, the HQC key generation function is substantially faster than SoA implementations.
Most importantly, the results show that our optimizations significantly reduce the performance gap between code-based and lattice-based algorithms, as shown by comparing the KYBER512 hardware implementation~\cite{xing2021compact} to our design.

\section{Conclusion}
In this work, we presented to the best of our knowledge the first in-depth study on the efficient implementation of the new, more secure HQC sampling procedure.
We introduced novel methodologies to implement the sampling algorithm that substantially enhances the efficiency of its hardware implementation and reduces the algorithmic complexity from $\mathcal{O}(n^2)$ to $\mathcal{O}(n)$.
The changes were carefully evaluated not to induce any new side channel attack possibilities or to comprise the security of the implementation.
Our results show that these improvements reduce the latency of the sampling algorithm by a factor of 24, while the hardware resources required remain comparable to those of the original sampling algorithm, particularly when implemented jointly with the ring arithmetic in HQC.
Based on these changes, and the fact that HQC was already shown to be the most efficient code-based cryptosystem in the NIST process, the performance gap between code-based and lattice-based algorithms substantially decreases - from a factor of 10.5 to a factor of 1.7 in terms of execution time - by integrating our proposals into a SoA hardware design of the HQC-128 key generation function.
To conclude, HQC is a very competitive alternative to KYBER in case it turns out to be insecure in the future in terms of resource efficiency.

\bibliographystyle{IEEEtran}
\bibliography{references}

\begin{thebibliography}{10}
\providecommand{\url}[1]{#1}
\csname url@samestyle\endcsname
\providecommand{\newblock}{\relax}
\providecommand{\bibinfo}[2]{#2}
\providecommand{\BIBentrySTDinterwordspacing}{\spaceskip=0pt\relax}
\providecommand{\BIBentryALTinterwordstretchfactor}{4}
\providecommand{\BIBentryALTinterwordspacing}{\spaceskip=\fontdimen2\font plus
\BIBentryALTinterwordstretchfactor\fontdimen3\font minus
  \fontdimen4\font\relax}
\providecommand{\BIBforeignlanguage}[2]{{%
\expandafter\ifx\csname l@#1\endcsname\relax
\typeout{** WARNING: IEEEtran.bst: No hyphenation pattern has been}%
\typeout{** loaded for the language `#1'. Using the pattern for}%
\typeout{** the default language instead.}%
\else
\language=\csname l@#1\endcsname
\fi
#2}}
\providecommand{\BIBdecl}{\relax}
\BIBdecl

\bibitem{mosca2018}
M.~Mosca, ``{Cybersecurity in an Era with Quantum Computers: Will We Be
  Ready?}'' \emph{IEEE Security \& Privacy}, vol.~16, no.~5, pp. 38--41, 2018.

\bibitem{nist2023}
{National Institute for Standards and Technology, U.S. Departement of
  Commerce}, ``{Post-Quantum Cryptography},'' 2023,
  \url{https://csrc.nist.gov/projects/post-quantum-cryptography}, 2023,
  Retrieved 2023-06-07.

\bibitem{melchor2021hamming}
C.~A. Melchor, N.~Aragon, S.~Bettaieb, L.~Bidoux, O.~Blazy, J.-C. Deneuville,
  P.~Gaborit, E.~Persichetti, G.~Z{\'e}mor, and I.~Bourges, ``{Hamming
  Quasi-Cyclic (HQC)},'' 2021.

\bibitem{aguilar2022towards}
C.~Aguilar-Melchor, J.-C. Deneuville, A.~Dion, J.~Howe, R.~Malmain,
  V.~Migliore, M.~Nawan, and K.~Nawaz, ``{Towards Automating Cryptographic
  Hardware Implementations: A Case Study of HQC},'' in \emph{Code-Based
  Cryptography Workshop}.\hskip 1em plus 0.5em minus 0.4em\relax Springer,
  2022, pp. 62--76.

\bibitem{schoffel2022code}
M.~Sch{\"o}ffel, J.~Feldmann, and N.~Wehn, ``{Code-based Cryptography in IoT: A
  HW/SW Co-Design of HQC},'' in \emph{2022 IEEE 8th World Forum on Internet of
  Things (WF-IoT)}.\hskip 1em plus 0.5em minus 0.4em\relax IEEE, 2022, pp.
  1--7.

\bibitem{tu2023leap}
Y.~Tu, P.~He, {\c{C}}.~K. Ko{\c{c}}, and J.~Xie, ``{LEAP: Lightweight and
  Efficient Accelerator for Sparse Polynomial Multiplication of HQC},''
  \emph{IEEE Transactions on Very Large Scale Integration (VLSI) Systems},
  2023.

\bibitem{krausz2023holistic}
M.~Krausz, G.~Land, J.~Richter-Brockmann, and T.~G{\"u}neysu, ``{A Holistic
  Approach Towards Side-Channel Secure Fixed-Weight Polynomial Sampling},'' in
  \emph{IACR International Conference on Public-Key Cryptography}.\hskip 1em
  plus 0.5em minus 0.4em\relax Springer, 2023, pp. 94--124.

\bibitem{sendrier2021secure}
N.~Sendrier, ``{Secure Sampling of Constant-Weight Words--Application to
  BIKE},'' \emph{Cryptology ePrint Archive}, 2021.

\bibitem{guo2022don}
Q.~Guo, C.~Hlauschek, T.~Johansson, N.~Lahr, A.~Nilsson, and R.~L.
  Schr{\"o}der, ``{Don’t reject this: Key-recovery timing attacks due to
  rejection-sampling in HQC and BIKE},'' \emph{IACR Transactions on
  Cryptographic Hardware and Embedded Systems}, pp. 223--263, 2022.

\bibitem{nist2015}
{National Institute of Standards and Technology}, ``{FIPS PUB 202 -SHA-3
  Standard: Permutation-Based Hash and Extendable-Output Functions},''
  \url{https://csrc.nist.gov/Projects/post-quantum-cryptography}, Retrieved
  2022-02-17.

\bibitem{becker2021neon}
H.~Becker, V.~Hwang, M.~J. Kannwischer, B.-Y. Yang, and S.-Y. Yang, ``{Neon
  NTT: Faster Dilithium, Kyber, and Saber on Cortex-A72 and Apple M1},''
  \emph{Cryptology ePrint Archive}, 2021.

\bibitem{bert2022}
{Bertoni, G. and Daemen, J. and Peeters, M. and Van Assche, G.}, ``{Keccak in
  VHDL},'' \url{https://keccak.team/hardware.html}, 2022, Retrieved 2023-06-07.

\bibitem{xing2021compact}
Y.~Xing and S.~Li, ``{A compact hardware implementation of CCA-secure key
  exchange mechanism CRYSTALS-KYBER on FPGA},'' \emph{IACR Transactions on
  Cryptographic Hardware and Embedded Systems}, pp. 328--356, 2021.

\end{thebibliography}

\end{document}